\documentstyle[twoside,fleqn,espcrc2,epsfig,amsfonts,latexsym]{article}

\newcommand{\AmS}{{\protect\the\textfont2
  A\kern-.1667em\lower.5ex\hbox{M}\kern-.125emS}}

\title{Finite Temperature LGT in a Finite Box
       with BPS Monopole Boundary Conditions
      \thanks{Talk given by. A.~I.~Veselov}}

\author{E.-M.~Ilgenfritz\address{Institute for Theoretical Physics,
			   Kanazawa University, Kanazawa 920-1192, Japan},
        S.~V.~Molodtsov\address{ITEP, B. Cheremushkinskaya 25,
			                           Moscow 117259, Russia},
        M.~M\"uller-Preussker\address{Institut f\"ur Physik,
			       Humboldt-Universit\"at zu Berlin, Germany},
	and A.~I.~Veselov$\mathrm{^b}$
	}

\begin{document}

\begin{abstract}
Finite temperature $SU(2)$ lattice gauge theory is investigated in a
$3D$ cubic box with fixed boundary conditions (b.c.) provided by a discretized,
static BPS monopole solution with varying core scale ${\mu}$.
For discrete ${\mu}$-values we find stable classical solutions either
of electro-magnetic ('dyon') or of purely magnetic type inside the box.
Near the deconfinement transition we study the influence of the
b.c. on the quantized fields inside the box. In contrast
to the purely magnetic background field case, for the dyon case
we observe confinement for temperatures above the usual critical one.
\end{abstract}

\maketitle

\section{INTRODUCTION}
Quark confinement in QCD is a long-standing problem which has not found
a satisfactory explanation until now. Several models have been considered.
The dual superconductor scenario \cite{mon} views confinement as a
dual Meissner effect due to the condensation of Abelian monopoles. But
it badly serves as a frame for phenomenological applications.
On the other hand
the instanton-based semi-classical approach has been very successful
in phenomenology. However, it has not yet explained confinement.
The question arises, whether other extended, maybe monopole-like, semi-classical
configurations may support confinement. In this talk we would like
to give a partial answer to this question.

The configuration we want to start from is the
Bogomol'nyi-Prasad-Sommerfield (BPS) mono\-pole  \cite{bps}
taken as a static solution of the equation of motion in Euclidean
$SU(2)$ Yang-Mills theory. Following an investigation reported by
Smit and van der Sijs \cite{smitvds1}
we discretize it on the lattice and keep the tangential boundary values
fixed for the 3-volume. With these b.c.
we find stable solutions
inside the box. Depending on the core scale there are configurations, which
can be either of purely magnetic type
-- t'Hooft-Polyakov-like (HP) monopoles \cite{thp} --
or of self-dual BPS-type.
The latter possess also electric charge. Therefore,
let us call them 'dyons'. With the stable solutions taken
as a well-defined semi-classical background we investigate Monte Carlo
generated quantum fields.

For the finite temperature case we shall see,
how usual indicators of the deconfinement phase transition
(average Polyakov line, distribution of Polyakov lines) behave
in the presence of the different background fields.

\section{CLASSICAL FIELDS WITH DYON BOUNDARY CONDITIONS}
Let us discuss classical field configurations on a hypercubic lattice
in a finite volume with fixed spatial b.c.
The latter will be specified such that link variables
being normal to the boundary and pointing outward are
inactive, {\it i.e.} cannot contribute to the
action. Link variables tangential to the boundaries will be fixed to
classical values given by a BPS dyon configuration.
For the imaginary time direction periodicity is implied.
We discretize the BPS dyon field as in \cite{smitvds1}.
Its center be fixed at the center of the $3D$ box.
The large distance asymptotic behaviour of the BPS dyon gauge potentials
provides a periodic behaviour of the boundary link variables as a function of
the core scale-size parameter $\mu$. It
allows to create b.c. compatible either with a BPS dyon
or with a HP monopole with suppressed electric fields.
Following \cite{smitvds1} it is convenient to parametrize
the boundary values by
an auxiliary parameter $\mu'$ instead of $~\mu$,
$~~\mu' = (\mu - 1 / R_{eff}) \cdot N_t / (2 \pi),~~$
where $~R_{eff} \simeq 1.13 \cdot (N_s-1)/2~$.

We wanted to find out, what lattice fields really
have minimal action for given b.c.
It turned out that there exist {\it several
local} minima of action for $~\mu'~$ sufficiently big.

In order to characterize the lattice field configurations obtained
by heating and cooling
we used the full Wilson plaquette action
$S = E^2 + B^2$,
with its magnetic part $B^2$ coming from the spatial plaquette contributions
and its electric part $E^2$  obtained from the time-like plaquettes.
\begin{figure}[!htb]
\vspace*{-1.0cm}
\begin{center}
\epsfig{file=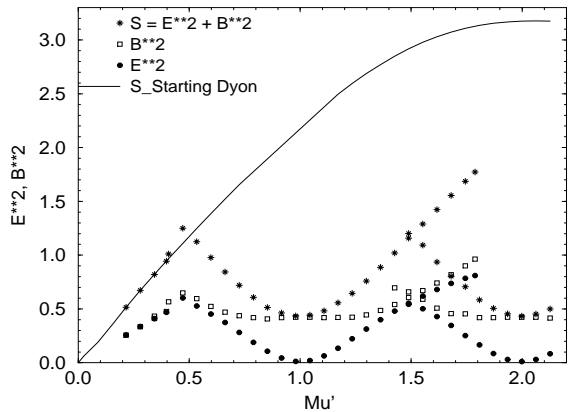,width=5.5cm,height=7.5cm,angle=270}
\end{center}
\vspace*{-1.0cm}
\caption{
Total action, magnetic and electric part of the action.
 }
\end{figure}
\vspace*{-0.2cm}

Fig. 1 shows the full action per time slice of the
classical continuum BPS dyon in dependence
on the parameter $\mu'$ (solid line). The action values are given in units of
$4 \pi / (a~g^2)$.
The data points in Fig. 1 show the global or local minima, respectively,
found for the full action (asterisks) after different attemps of heating
and subsequent cooling. The most interesting for us are the states
obtained for half-integer and integer $~\mu'$--values, respectively.
For $ \mu' \simeq 0.5, 1.5, \cdots $ we reproduce a BPS dyon state
identical to the original one at $ \mu' = 0.5 $ with equal
electric and magnetic contributions to the action.
For $ \mu' \simeq 1.0, 2.0, \cdots $ only the magnetic part of
the action survives. That means that we have obtained purely magnetic,
HP-like monopoles.

\section{QUANTUM FIELDS WITH DYON BOUNDARY CONDITIONS}
In the following we want to investigate quantum fields
for well-defined background fields
being either BPS dyon or HP monopole classical lattice solutions
depending on the spatial b.c.
The simulations have been done with a standard Metropolis algorithm
on a $16^3 \times 4$ lattice. For $N_t=4$ the
deconfinement phase transition occurs at
$~\beta = \beta_c \simeq 2.29$ in the large-volume limit.
In the following we will mainly consider two typical cases:
$\beta=2.2$, characteristic for the confinement phase, and $\beta=2.4$,
for the deconfinement phase.

By computing space-like and time-like average plaquettes as a function
of the distance $d$ from the spatial boundary
one can define a core inside the three-volume ($d \ge 3$),
where these local observables take on a plateau value.

We have computed the average Polyakov line as a function of the distance $d$
from the boundary. Even inside the core (at $d \geq 3$) the behaviour of the
Polyakov line strongly depends on the b.c., {\it i.e.} whether there is
a BPS dyon or a HP monopole background field. For integer
$\mu'$ (HP monopole) in both the confinement and the deconfinement phase
the average Polyakov line approaches the value characteristic for
this $\beta$ (with periodic b.c.).
On the contrary, for half-integer $\mu'$ (BPS dyon) the average
Polyakov line remains  very small throughout the whole lattice when
passing the deconfinement transition.
At $\beta = 2.4$ inside the core we find $|<L>| < 0.1$.

These observations are supported by the corresponding histograms of local
Polyakov line values measured at distance $d=5$ from the boundary
which are shown in Fig. 2 for
$\beta = 2.2$ and $2.4$, respectively.
\begin{figure}[!htb]
\begin{center}
\begin{tabular}{cc}
$\beta = 2.2$ \\
\epsfig{file=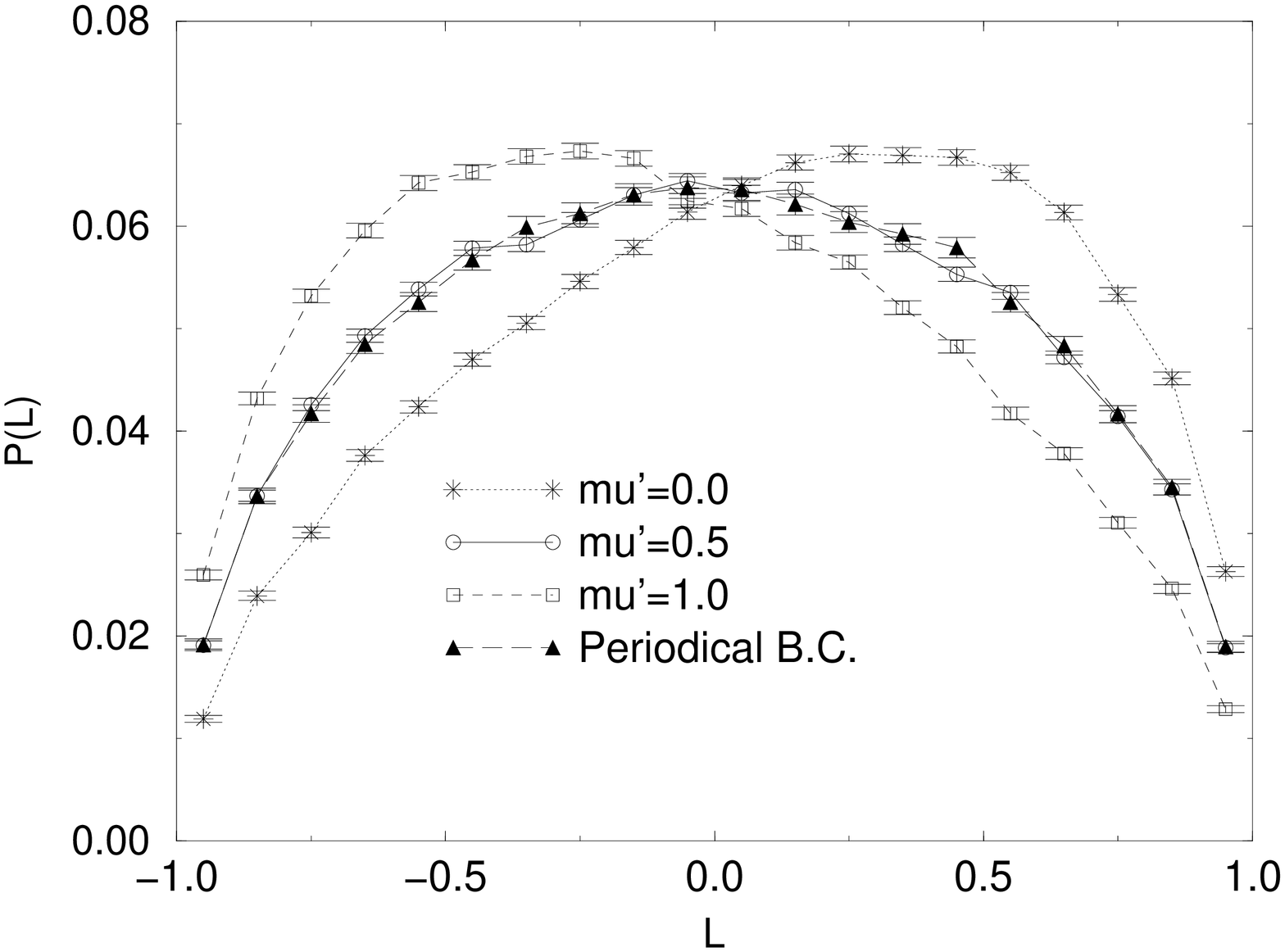,width=5.4cm,height=7.0cm,angle=270}\\
\\
$\beta = 2.4$ \\
\epsfig{file=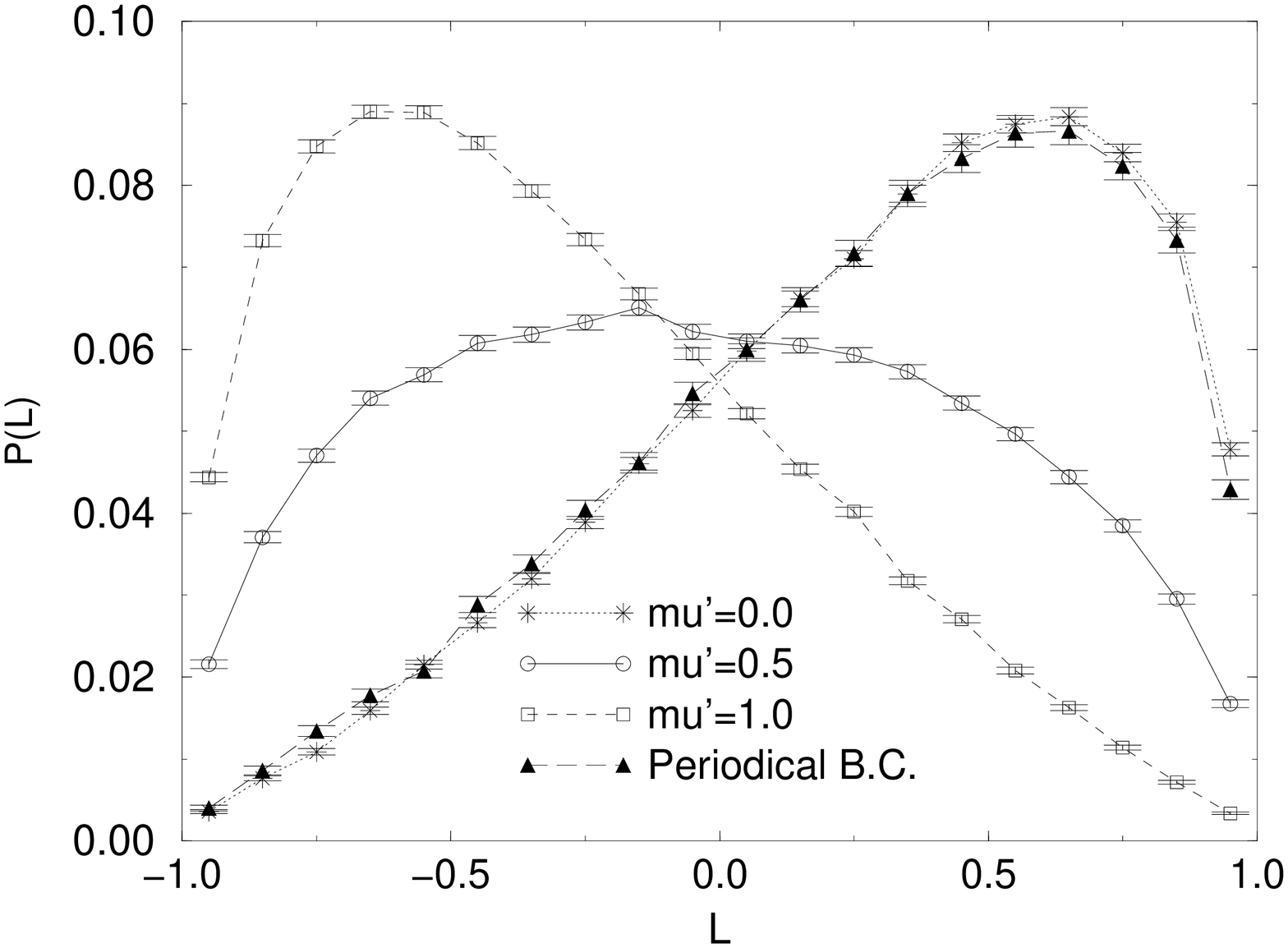,width=5.4cm,height=7.0cm,angle=270}\\
\end{tabular}
\end{center}
\vspace*{-1.0cm}
\caption{
Distributions of local Polyakov line values for various b.c.
 }
\end{figure}
%
In the confinement phase, the histograms corresponding to the monopole
b.c. are displaced representing  entirely the effect of the finite
correlation length of the Polyakov line, 
while the histogram for the dyonic
b.c. coincides with that for periodic b.c.
In the deconfinement phase the ''normal'' histogram (generated by staying
within one of the possible $Z(2)$-states with periodic b.c.)
is reproduced with monopole b.c. while the histogram for dyonic b.c.
is almost symmetric, similar to the lower $\beta$-value.

We conclude that, while the HP monopole b.c. are
compatible also with the confinement phase, a BPS dyon background
inside the finite box keeps the system
in the confinement state even for temperatures above $T_c$.
But as we have seen for smaller $N_t$ or larger $\beta$,
this does not persist at arbitrarily high temperature.

\section{CONCLUSIONS}
Our main observation was that in the dyon background field case
the quantum system inside a finite $3D$ box is kept in the confinement
state even for higher temperatures,
where for periodic boundary conditions deconfinement is caused.
The Polyakov line as the main order parameter for the
deconfinement transition turned out to be zero remarkably stable
inside the whole $3D$ volume considered.  However, a further increase of the
temperature finally restores deconfinement inside the $3D$  box.
The phenomenon reported here might be called {\it evaporation delay of
confinement} due to dyon boundary conditions.

We would like to interprete our findings in more general terms
that the confinement phenomenon is
strongly related to self-dual semi-classical objects rather than to
purely magnetic background fields.
On the other hand, deconfinement, when described semi-classically, requires
background fields which are not (anti-)self-dual.

\section*{ACKNOWLEDGEMENTS}
The authors are grateful to P. van Baal, B.V. Martemyanov, M.I. Polikarpov,
A.J. van der Sijs, Yu.A. Simonov, and J. Smit for useful discussions.
This work was partly supported by RFBR grants 97-02-17491 and 96-02-17230,
INTAS-RFBR 95-0681  and INTAS 96-370
and the joint DFG-RFFI grant 436 RUS 113/309/0 (R) or
RFBR-DFG grant 96-02-0088G.

\end{document}